\documentclass[aps,showpacs,superscriptaddress,preprint]{revtex4}%
\usepackage{mathrsfs}
\usepackage{amsfonts}
\usepackage{amsmath}
\usepackage{amssymb}
\usepackage{graphicx}
\usepackage{color}%
\begin{document}
\title{Multiple scattering theory of quasiparticles on a topological insulator surface}
\author{Zhen-Guo Fu}
\affiliation{SKLSM, Institute of Semiconductors, Chinese Academy of Sciences, P. O. Box
912, Beijing 100083, People's Republic of China}
\affiliation{LCP, Institute of Applied Physics and Computational Mathematics, P.O. Box
8009, Beijing 100088, People's Republic of China}
\author{Ping Zhang}
\thanks{zhang\_ping@iapcm.ac.cn}
\affiliation{LCP, Institute of Applied Physics and Computational Mathematics, P.O. Box
8009, Beijing 100088, People's Republic of China}
\author{Zhigang Wang}
\affiliation{LCP, Institute of Applied Physics and Computational Mathematics, P.O. Box
8009, Beijing 100088, People's Republic of China}
\author{Shu-Shen Li}
\thanks{sslee@semi.ac.cn}
\affiliation{SKLSM, Institute of Semiconductors, Chinese Academy of Sciences, P. O. Box
912, Beijing 100083, People's Republic of China}

\begin{abstract}
A general partial-wave multiple scattering theory for scattering from
cylindrically symmetric potentials on a topological insulator (TI) surface is
developed. As an application, the cross sections for a single scatterer and
two scatterers are discussed. We find that the symmetry of differential cross
section is reduced and the backscattering is allowed for massive Dirac
fermions on gapped TI surface. Remarkably, a sharp resonance peak at the band
edge of the gapped TI is found in the total cross section $\Lambda_{tot}$,
which may offer a useful way to determine the gap (as well as the effective
mass of quasiparticles) on TI surface. We show that the interference effect is
obvious in cross sections during the quasiparticle scattering between the
scatterer pair, and additional resonance peaks are introduced in
$\Lambda_{tot}$ when the higher partial waves are taken into account.

\end{abstract}

\maketitle


Topological insulator (TI) has attracted lots of interest in modern
condensed matter physics field since the advanced progress in
theoretical \cite{Kane,Kane1, Bernevig,Fu,Fang} and experimental
\cite{Konig,Hsieh1,Chen, Xia} studies of this new quantum matter
phase. The ideal three-dimensional (3D) TI exhibits gapless modes on
surface, and the backscattering is forbidden due to the time
reversal invariance, which has been observed in the angle-resolved
photo emission spectroscopy experiments on Bi$_{2}$Te$_{3}$ surface
\cite{Hsieh2}. However, if the Dirac fermions gain mass, a gap will
be induced in system, which breaks the time reversal symmetry
\cite{Qi, Hasan,Maciejko} and as a consequence, the scattering and
transport properties should be different from the gapless case. Due
to the single Dirac-cone nature, no complicated intervalley
scattering events occur on TI surface. Therefore, it can be expected
that \textit{in situ} measurement or even manipulation of impurity
scattering processes on TI surfaces should pave a promising way to
study electronic structures and find extraordinary quantum phenomena
in various TI materials. Recently, a series of scanning tunneling
microscopy (STM) measurements \cite{Xue, Roushan, Gomes} and
theoretical simulations \cite{Liu, Lee} have been performed on
Bi$_{2}$Te$_{3}$ and Bi$_{1-x}$Sb$_{x}$ surface states to study the
scattering effect in the dilute impurity limit. On the other hand,
when the impurities are located close to each other, the multiple
scattering effects should be important. Especially, since the
quasipartcle's spin is strongly coupled to its momentum, then the
quantum interference between different spin states during multiple
scattering process may display profound phenomena such as electric
conductance weak (anti-)localization \cite{He} and Aharonov-Bohm
effect \cite{Fu1} in STM signal. Therefore, it is clear that at
present more deep multiple scattering studies are being called for.

In this work we conduct a multiple scattering theory for the massive as well
as massless Dirac fermions on surfaces of 3D TIs. A general partial-wave
multiple scattering formula for scattering from short-range cylindrically
symmetric potentials is developed. The cross sections for a single scatterer
and two scatterers are discussed, and the interference effect during the
quasiparticle scattering between the scatterer pair is found. We show that
higher partial waves induce prominent corrections in cross sections, including
additional resonance peaks. The symmetry of differential cross section is
reduced and the backscattering can be observed for massive Dirac fermions due
to the breaking of the time reversal symmetry. Especially, a sharp resonance
peak at the band edge of the gapped Dirac fermion is found in total cross
section, which may offer a useful way to determine the gap of the TI surface
states. Note that our formulation is closely related to those given on
semiconductor heterostructure \cite{Walls1} and graphene
\cite{Novikov,Braun,Walls}. Remarkably, unlike graphene, the TI surface is a
single Dirac-cone system, which makes the present intravalley multiple
scattering theory more natural and exact.


We start from the effective-mass Hamiltonian near the Dirac point,%
\begin{equation}
H_{0}=\hbar v_{F}\left(  \boldsymbol{\sigma}\times\boldsymbol{k}\right)
\cdot\hat{z}+\Delta\sigma_{3},\label{1}%
\end{equation}
where $v_{F}\mathtt{\approx}5.0\mathtt{\times}10^{5}$m/s is the Fermi
velocity, $\boldsymbol{\sigma}\mathtt{=}\left(  \sigma_{1},\sigma_{2}%
,\sigma_{3}\right)  $ are Pauli matrices, $\boldsymbol{k}\mathtt{=}\left(
k_{x},k_{y}\right)  $ is the planer momentum, and $\Delta\mathtt{=}m^{\ast
}v_{F}^{2}$ is the band-gap, which is absent in the massless limit. The
eigenstates of $H_{0}$ are given by plane wave
\begin{equation}
\Psi_{0,\pm}\left(  \boldsymbol{r}\right)  =\frac{e^{i\boldsymbol{k}%
\cdot\boldsymbol{r}}}{\sqrt{2\left\vert \epsilon\right\vert }}\left(
\begin{array}
[c]{cc}%
\delta_{+}, & \mp i\delta_{-}e^{i\theta_{\boldsymbol{k}}}%
\end{array}
\right)  ^{T},\label{2}%
\end{equation}
where $\delta_{\pm}\mathtt{=}\sqrt{\left\vert \epsilon\pm\Delta\right\vert },$
and $\epsilon\mathtt{=}\pm\sqrt{\left(  \hbar v_{F}k\right)  ^{2}%
\mathtt{+}\Delta^{2}}$ with the upper (lower) sign referring to the electron
(hole) part of the spectrum. The Berry phase of system given by $-i\int
_{0}^{2\pi}d\theta_{k}\left\langle \Psi_{0,\pm}\right\vert \partial
_{\theta_{k}}\left\vert \Psi_{0,\pm}\right\rangle \mathtt{=}\frac{\left\vert
\epsilon-\Delta\right\vert }{\left\vert \epsilon\right\vert }\pi$ indicates
that the backscattering is allowed (prohibited) if band-gap $\Delta
\mathtt{\neq}0$ ($\mathtt{=}0$), which will be observed in differential cross
section in the following discussion. To obtain the scattering theory from
localized, cylindrically symmetric scatterers, it is convenient for boundary
condition treatment to express the eigenstates for Eq. (\ref{1}) in
cylindrical coordinates, which are written as%
\begin{equation}
\chi_{l,\pm}^{(1,2)}\left(  \boldsymbol{r}\right)  =\sqrt{\frac{k}{2\left\vert
\epsilon\right\vert }}e^{il\theta}\left(
\begin{array}
[c]{cc}%
\delta_{+}H_{l}^{(1,2)}, & \pm\delta_{-}H_{l+1}^{(1,2)}e^{i\theta}%
\end{array}
\right)  ^{T},\label{3}%
\end{equation}
for outgoing and incoming cylindrical waves about $\boldsymbol{r}$=$0$,
respectively. Here $H_{l}^{(1,2)}$ are Hankel functions of order $l$ with
variable $kr$.

The potential of each symmetric scalar scatterer can be expressed as%
\begin{equation}
V\left(  \boldsymbol{r}\right)  =V_{0}\theta\left(  a-r\right)  ,\label{4}%
\end{equation}
where $a$ is the radius of scatterer. The incident plane wave centered about a
single scatterer located at $\boldsymbol{r}_{n}$ is given by%
\begin{equation}
\Phi_{\pm}^{in}\left(  \boldsymbol{r}\right)  =\frac{e^{i\boldsymbol{k}%
\cdot\boldsymbol{r}_{n}}}{2\sqrt{k}}\sum_{l=-\infty}^{\infty}e^{il\left(
\theta_{n}-\theta_{\boldsymbol{k}}\right)  }i^{l}\left(  \chi_{l,\pm}%
^{(1)}\left(  \boldsymbol{\rho}_{n}\right)  +\chi_{l,\pm}^{(2)}\left(
\boldsymbol{\rho}_{n}\right)  \right)  ,\label{5}%
\end{equation}
where $\chi_{l,\pm}^{(1)}$ and $\chi_{l,\pm}^{(2)}$ represent outgoing and
incoming waves about $\boldsymbol{r}_{n}$, with $\boldsymbol{\rho}%
_{n}\mathtt{=}\boldsymbol{r}\mathtt{-}\boldsymbol{r}_{n}$ and $e^{i\theta_{n}%
}\mathtt{=}\frac{\boldsymbol{\rho}_{n}\cdot\left(  \hat{x}+i\hat{y}\right)
}{\rho_{n}}$. By using the boundary condition at $\rho_{n}\mathtt{=}a$, we can
get the scattered wave function%
\begin{equation}
\Psi_{\pm}^{sc}\left(  \boldsymbol{r}\right)  =s_{0}G_{+0}T_{0}^{-}\Phi_{\pm
}^{in}+\sum_{l=1}^{\infty}\left[  s_{l}G_{+l}T_{l}^{-}+s_{-l}G_{-l}T_{l}%
^{+}\right]  \Phi_{\pm}^{in},\label{e6}%
\end{equation}
where the $l$th\texttt{-}partial wave $t\mathtt{-}$matrix is $T_{l}^{\pm
}\mathtt{=}$diag$(%
\begin{array}
[c]{cc}%
\hat{P}_{l}^{\pm}, & \mp i\hat{P}_{l\mp1}^{\pm}%
\end{array}
)$ with $\hat{P}_{l}^{\pm}\mathtt{=}\frac{1}{i^{l}k^{l}}(\partial
_{x}\mathtt{\pm}i\partial_{y})^{l}\mathtt{=}\frac{e^{\pm il\theta}}{i^{l}%
k^{l}}(\partial_{r}\mathtt{\pm}\frac{i}{r}\partial_{\theta})^{l}$, and
\begin{align}
G_{+l} &  =\frac{i^{l}e^{il\theta_{n}}}{2\left\vert \epsilon\right\vert
}\left(
\begin{array}
[c]{cc}%
\delta_{+}^{2}H_{l}^{(1)} & \pm\delta_{+}\delta_{-}H_{l}^{(1)}\\
\pm\delta_{+}\delta_{-}H_{l+1}^{(1)}e^{i\theta_{n}} & \delta_{-}^{2}%
H_{l+1}^{(1)}e^{i\theta_{n}}%
\end{array}
\right)  ,\nonumber\\
G_{-l} &  =\frac{i^{l}e^{-il\theta_{n}}}{2\left\vert \epsilon\right\vert
}\left(
\begin{array}
[c]{cc}%
\delta_{+}^{2}H_{l}^{(1)} & \mp\delta_{+}\delta_{-}H_{l}^{(1)}\\
\mp\delta_{+}\delta_{-}H_{l-1}^{(1)}e^{i\theta_{n}} & \delta_{-}^{2}%
H_{l-1}^{(1)}e^{i\theta_{n}}%
\end{array}
\right)  .\label{7-1}%
\end{align}
The variable of Hankel functions is $k\rho_{n}$. The scattering amplitude of
the $l$th partial wave is
\begin{equation}
s_{l}=\frac{\tilde{A}^{+}J_{l}\left(  ka\right)  J_{l+1}\left(  k^{\prime
}a\right)  -\tilde{A}^{-}J_{l+1}\left(  ka\right)  J_{l}\left(  k^{\prime
}a\right)  }{\tilde{A}^{-}H_{l+1}^{(1)}\left(  ka\right)  J_{l}\left(
k^{\prime}a\right)  -\tilde{A}^{+}H_{l}^{(1)}\left(  ka\right)  J_{l+1}\left(
k^{\prime}a\right)  },\label{7}%
\end{equation}
where $\tilde{A}^{\pm}\mathtt{=}\sqrt{\left\vert \frac{\epsilon\pm\Delta
}{\epsilon-V_{0.}\pm\Delta}\right\vert }$, $k^{\prime}\mathtt{=}\sqrt{\left(
\epsilon\mathtt{-}V_{0}\right)  ^{2}\mathtt{-}\Delta^{2}}/\hbar v_{F}$, and
$J_{l}$ is the Bessel function of order $l$. In the massless limit, i.e.,
$\Delta\mathtt{=}0$, the above Eq. (\ref{e6}) can be reduced into a more
compact form due to the symmetry of the band, which is written as%
\begin{equation}
\Psi_{\pm}^{sc}\left(  \boldsymbol{r}\right)  =\sum_{l=0}^{\infty}%
\frac{4i\hbar v_{F}s_{l}}{k}G_{l}\left(  \boldsymbol{r},\boldsymbol{r}%
_{n},\epsilon\right)  T_{l}\left[  \Phi_{\pm}^{in}\right]  ,\label{8}%
\end{equation}
where
\begin{equation}
G_{l}\propto\left(
\begin{array}
[c]{cc}%
H_{l}^{(1)}e^{il\theta_{n}} & \mp H_{l+1}^{(1)}e^{-i\left(  l+1\right)
\theta_{n}}\\
\pm H_{l+1}^{(1)}e^{i\left(  l+1\right)  \theta_{n}} & H_{l}^{(1)}%
e^{-il\theta_{n}}%
\end{array}
\right)  ,\label{9}%
\end{equation}
and $T_{l}\mathtt{=}\mathtt{diag}(%
\begin{array}
[c]{cc}%
\hat{P}_{l}^{-}, & \hat{P}_{l}^{+}%
\end{array}
)$. Here, we have used the relation $s_{-(l+1)}\mathtt{=}s_{l}$ for
$\Delta\mathtt{=}0$.

It is easy to extend the above theory to multiple scattering
problems of massive quasiparticles on TI surface. The total wave
function for $N$
scatterers located at positions $\boldsymbol{r}_{1},\boldsymbol{r}_{2}%
,\cdots\boldsymbol{r}_{N}$ is given by $\Psi\left(  \boldsymbol{r}\right)
\mathtt{=}\Phi_{\pm}^{in}\mathtt{+}\Psi_{\pm}^{sc}$, where
\begin{equation}
\Psi_{\pm}^{sc}\left(  \boldsymbol{r}\right)  =\mathbb{G}\left(
\boldsymbol{r}\right)  \mathit{S}M^{-1}\vec{\phi}.\label{10}%
\end{equation}
Here, $\mathbb{G}\left(  \boldsymbol{r}\right)  $ is a $2\mathtt{\times
}2N\left(  2l_{\max}\mathtt{+}1\right)  $ matrix,
\begin{equation}
\mathbb{G}\left(  \boldsymbol{r}\right)  =\left(
\begin{array}
[c]{ccc}%
\tilde{G}\left(  \boldsymbol{r},\boldsymbol{r}_{1}\right)  , & \tilde
{G}\left(  \boldsymbol{r},\boldsymbol{r}_{2}\right)  , & \cdots,
\end{array}%
\begin{array}
[c]{c}%
\tilde{G}\left(  \boldsymbol{r},\boldsymbol{r}_{N}\right)
\end{array}
\right)  \label{11}%
\end{equation}
with $\tilde{G}\mathtt{=}[%
\begin{array}
[c]{ccc}%
G_{+0}, & G_{+1}, & G_{-1},
\end{array}
\cdots,%
\begin{array}
[c]{cc}%
G_{+l_{\max}}, & G_{-l_{\max}}%
\end{array}
]$. $\mathit{S}$ is a $2N\left(  2l_{\max}\mathtt{+}1\right)  \mathtt{\times
}2N\left(  2l_{\max}\mathtt{+}1\right)  $ diagonal matrix with the nontrivial
elements $s_{0}$ and $s_{\pm l}$. $M$ is expressed by%
\begin{equation}
M=\mathbf{1}-\mathbf{G}\mathit{S},\label{13}%
\end{equation}
where
\begin{equation}
\mathbf{G=}\text{$\left(
\begin{array}
[c]{cccc}%
\mathbf{0} & \mathbf{G}\left(  1,2\right)   & \mathbf{\cdots} & \mathbf{G}%
\left(  1,N\right)  \\
\mathbf{G}\left(  2,1\right)   & \mathbf{0} & \mathbf{\cdots} & \mathbf{G}%
\left(  2,N\right)  \\
\mathbf{\vdots} & \mathbf{\vdots} & \mathbf{\ddots} & \vdots\\
\mathbf{G}\left(  N,1\right)   & \mathbf{G}\left(  N,2\right)   & \cdots &
\mathbf{0}%
\end{array}
\right)  ,$}\label{14}%
\end{equation}
with $\mathbf{G}\left(  i,j\right)  $ are $2\left(  2l_{\max}\mathtt{+}%
1\right)  \mathtt{\times}2\left(  2l_{\max}\mathtt{+}1\right)  $ matrices
constructed by $T_{l}^{\pm}\left[  G_{l^{\prime}}^{k}\left(  i,j\right)
\right]  $ ($0\mathtt{\leq}l,l^{\prime}\mathtt{\leq}l_{\max}$). Finally,
$\vec{\phi}$ can be written as a $2N\left(  2l_{\max}\mathtt{+}1\right)
\mathtt{\times}1$ vector, explicitly,
\begin{equation}
\vec{\phi}=\left(
\begin{array}
[c]{cccc}%
\phi_{1}, & \phi_{2}, & \cdots, & \phi_{N}%
\end{array}
\right)  ^{T},\label{15}%
\end{equation}
where $\phi_{i}\mathtt{=}\left[
\begin{array}
[c]{cccc}%
T_{0}^{-}\left[  \Phi\left(  \boldsymbol{r}_{i}\right)  \right]  , & T_{1}%
^{-}\left[  \Phi\left(  \boldsymbol{r}_{i}\right)  \right]  , & T_{1}%
^{+}\left[  \Phi\left(  \boldsymbol{r}_{i}\right)  \right]  , & \mathtt{\cdots
},
\end{array}
\right.  \left.
\begin{array}
[c]{cc}%
T_{l_{\max}}^{-}\left[  \Phi\left(  \boldsymbol{r}_{i}\right)  \right]  , &
T_{l_{\max}}^{+}\left[  \Phi\left(  \boldsymbol{r}_{i}\right)  \right]
\end{array}
\right]  ^{T}$. The multiple scattering problem will be simplified in the
massless limit $\Delta\mathtt{=}0$, and similar to the case of a single
scatterer, some transformations, including $s_{-(l+1)}\mathtt{=}s_{l}$,
$(T_{l}^{+},T_{l}^{-})\mathtt{\rightarrow}T_{l}$, and $(G_{+l},G_{-l}%
)\mathtt{\rightarrow}G_{l}$, should be performed in Eqs. (\ref{10}-\ref{15}).
Consequently, matrices $M$ and $\mathit{S}$ are reduced to $2N\left(  l_{\max
}\mathtt{+}1\right)  \mathtt{\times}2N\left(  l_{\max}\mathtt{+}1\right)  $
ones, which are similar to the problems of intravalley multiple scattering of
quasiparticles in graphene \cite{Walls}, so that we do not show the explicit
formulas for the case of $\Delta\mathtt{=}0$ herein.


The above theory allows one to solve the multiple scattering problems on
gapped or gapless TI surface with higher partial waves, which may be important
as the distance between scatterers decreases as well as the scattering
potential is strong. If the potential is very weak, $s\mathtt{-}$wave
($l_{\max}$=$0$) is enough. To obtain the scattering amplitude and the cross
section, we apply the following approximation on Hankel functions
$H_{l}^{\left(  1\right)  }$ in Eq. (\ref{11}) for $r\mathtt{\rightarrow
}\infty$,
\begin{equation}
H_{l}^{\left(  1\right)  }\left(  k\rho_{n}\right)  \rightarrow\sqrt{\frac
{2}{i^{2l+1}\pi kr}}e^{ikr}e^{-ik\hat{r}\cdot\boldsymbol{r}_{n}},\label{16}%
\end{equation}
and $e^{i\theta_{n}}\mathtt{=}\boldsymbol{\rho}_{n}\mathtt{\cdot}\left(
\hat{x}\mathtt{+}i\hat{y}\right)  /\rho_{n}\mathtt{\rightarrow}\hat
{r}\mathtt{\cdot}\left(  \hat{x}\mathtt{+}i\hat{y}\right)  \mathtt{=}%
e^{i\varphi}$, where $\hat{r}\mathtt{=}\left(  \cos\varphi\mathtt{,}%
\sin\varphi\right)  $ is a unit vector in the direction of $\boldsymbol{r}$.
Finally, the scattered wave function can be approximated as
\begin{equation}
\Psi_{\pm}^{sc}\left(  \boldsymbol{r}\right)  \rightarrow f\left(
\boldsymbol{k},\varphi\right)  \frac{1}{\sqrt{2\left\vert \epsilon\right\vert
}}\left(
\begin{array}
[c]{cc}%
\delta_{+}, & \mp i\delta_{-}e^{i\varphi}%
\end{array}
\right)  ^{T}\frac{e^{ikr}}{\sqrt{r}},\label{17}%
\end{equation}
where $f\left(  \boldsymbol{k},\varphi\right)  $ is nothing else but the
scattering amplitude. Therefore, the differential cross section should be
\begin{equation}
d\Lambda/d\varphi=\left\vert f\left(  \boldsymbol{k},\varphi\right)
\right\vert ^{2},
\end{equation}
and the total cross section is
\begin{equation}
\Lambda_{tot}=\int_{0}^{2\pi}d\varphi\left\vert f\left(  \boldsymbol{k}%
,\varphi\right)  \right\vert ^{2}=C_{\Delta}\sqrt{\frac{8\pi}{k}%
}\operatorname{Im}\left[  e^{-i\frac{\pi}{4}}f\left(  \boldsymbol{k}%
,\varphi=0\right)  \right]  ,\label{20}%
\end{equation}
The second equality for $\Lambda_{tot}$ expresses the 2D optical theorem in
terms of our definition of $f$. $C_{\Delta}\mathtt{=}\left(  \delta_{+}%
^{2}\mathtt{+}\delta_{-}^{2}\right)  /2\left\vert \epsilon\right\vert $ arises
from the band gap, which reduces to unity in the massless limit. On the other
hand, the scattered current $\boldsymbol{j}_{sc}\mathtt{=}\mathtt{\mp}%
v_{F}\frac{\delta_{+}\delta_{-}}{\left\vert \epsilon\right\vert }%
\frac{\left\vert f\left(  \boldsymbol{k},\varphi\right)  \right\vert ^{2}}%
{r}\left[  \cos\varphi\hat{x}\mathtt{+}\sin\varphi\hat{y}\right]  $, and the
incident current $\boldsymbol{j}_{in}\mathtt{=}\mathtt{\mp}v_{F}\frac
{\delta_{+}\delta_{-}}{\left\vert \epsilon\right\vert }\left[  \cos\theta
_{k}\hat{x}\mathtt{+}\sin\theta_{k}\hat{y}\right]  $, which result in
$d\Lambda/d\varphi\mathtt{=}r\left\vert \boldsymbol{j}_{sc}\right\vert
/\left\vert \boldsymbol{j}_{in}\right\vert \mathtt{=}\left\vert f\left(
\boldsymbol{k},\varphi\right)  \right\vert ^{2}$. In the following
calculations, without losing generality, we just consider the incident wave
$\Phi_{+}^{in}\mathtt{=}\frac{e^{ikx}}{\sqrt{2\left\vert \epsilon\right\vert
}}(%
\begin{array}
[c]{cc}%
\delta_{+}\mathtt{,} & \mathtt{-}i\delta_{-}%
\end{array}
)^{T}$ propagating along the positive $\hat{x}$ direction. \begin{figure}[ptb]
\begin{center}
\includegraphics[width=0.6\linewidth]{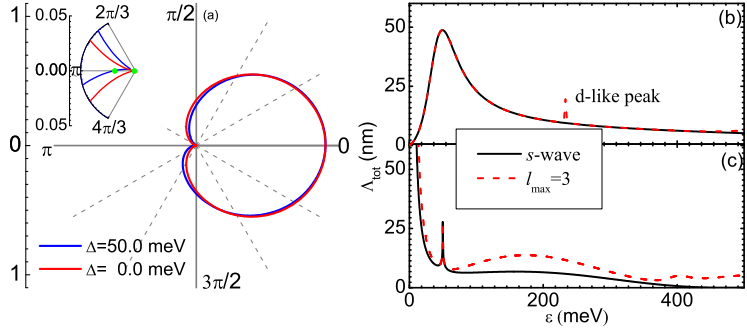}
\end{center}
\caption{ (Color online) The normalized differential (a) and the total (b-c)
cross sections for a single scatterer. In (a), the red (blue) curve is for
$\Delta\mathtt{=}0$ meV ($\Delta\mathtt{=}50$ meV), and $\epsilon
\mathtt{=}320$ meV. (b) is for $\Delta\mathtt{=}0$ meV, while (c) is for
$\Delta\mathtt{=}50$ meV. Insert in (a) is a zoom of $d\Lambda/d\varphi$ in
angle range $[2\pi/3,4\pi/3]$, and the green dots indicate that the
backscattering is allowed (forbidden) in the case of $\Delta\mathtt{\neq}0$
($\Delta\mathtt{=}0$). In all calculations, $V_{0}\mathtt{=}2$ eV, and
$a\mathtt{=}1$ nm are chosen. }%
\label{fig1}%
\end{figure}


Firstly, we discuss the single nonmagnetic scatterer problem. The numerical
results of scattering cross section are shown in Fig. \ref{fig1}. From Fig.
\ref{fig1}(a), one can observe that the backscattering at potential is absent
when the band gap decreases to zero, i.e., the differential cross section
$\left.  d\Lambda\left(  \Delta\mathtt{=}0\right)  /d\varphi\right\vert
_{\varphi=\pi}\mathtt{=}0$ [see the red curve in Fig. \ref{fig1}(a)]. This is
also can be seen from the $l$th$\mathtt{-}$partial wave scattering amplitude
\begin{equation}
f_{l}^{\Delta=0}=2\sqrt{\frac{2}{i\pi k}}s_{l}\cos\left[  \left(  2l+1\right)
\varphi/2\right]  e^{-i\varphi/2},\label{21}%
\end{equation}
which results in $f_{l}^{\Delta=0}\left(  \varphi\mathtt{=}\pi\right)
\mathtt{=}0$. However, on the band-gapped TI surface, the $l$th partial wave
scattering amplitude becomes
\begin{equation}
f_{\pm l}^{\Delta\neq0}=C_{\Delta}\sqrt{\frac{2}{i\pi k}}s_{\pm l}e^{\pm
il\varphi},\label{22}%
\end{equation}
which indicates $\left.  d\Lambda\left(  \Delta\mathtt{\neq}0\right)
/d\varphi\right\vert _{\varphi=\pi}\mathtt{=}\left\vert \sum_{l=0}^{l_{\max}%
}f_{\pm l}^{\Delta\neq0}\left(  \varphi\mathtt{=}\pi\right)  \right\vert
^{2}\mathtt{\neq}0$ since the time-reversal symmetry is broken by the nonzero
gap in the surface Dirac spectrum. The blue curve in Fig. \ref{fig1}(a) is for
$d\Lambda/d\varphi$ corresponding to $\Delta\mathtt{=}50$ meV. It is clear
that the symmetry of $d\Lambda\left(  \Delta\mathtt{\neq}0\right)  /d\varphi$
is reduced and the backscattering is allowed. This can be more clearly seen
from the insert in \ref{fig1}(a), which enlarges $d\Lambda/d\varphi$ in angle
range $\left[  2\pi/3,4\pi/3\right]  $.

The total cross section $\Lambda_{tot}$ as a function of energy $\epsilon$ is
shown in Fig. \ref{fig1}(b) and \ref{fig1}(c) for the gapless and gapped TI
surface. On average, the contributions from higher partial waves are much
smaller than $s\mathtt{-}$wave ($\left\vert f_{0}\right\vert \mathtt{\gg
}\left\vert f_{l\neq0}\right\vert $) in low-energy scattering regime, see the
black curves in Fig. \ref{fig1}(b) and \ref{fig1}(c) which only takes into
account the $s\mathtt{-}$wave in calculations. However, we find that higher
partial waves may introduce some remarkable corrections. For example, we find
an additional resonance peak ($d\mathtt{-}$like peak) around $\epsilon
\mathtt{=}233$ meV when the $f_{l=2}^{\Delta=0}$ partial wave is taken into
account, while much higher partial wave has little corrections on
$\Lambda_{tot}$, see the red dashed curves in Fig. \ref{fig1}(b). This
additional peak arises from the pole of $\left\vert s_{l=2}\right\vert $,
which could be obtained by expanding $\left\vert s_{l=2}\right\vert $ at this
point. The pole of $\left\vert s_{l=1}\right\vert $ is out of the energy range
we considered here, therefore, the corrections from $p\mathtt{-}$wave is small
in Fig. \ref{fig1}(b). Furthermore, we can find that $\Lambda_{tot}$ for
$\Delta\mathtt{\neq}0$ [Fig. \ref{fig1}(c)] is totally different from that for
$\Delta\mathtt{=}0$. A sharp peak at the band edge $\epsilon$=$\Delta$ is
found. In fact, $\epsilon$=$\Delta$ is a pole of $\left\vert s_{l}\right\vert
$ for arbitrary $l$th partial wave. Therefore, we can conclude that this sharp
peak in $\Lambda_{tot}$, which can be easily measured in scattering
experiment, may provide a useful way to determine the band gap (as well as the
effective mass $m^{\ast}$ of quasiparticle) of TI surface. The contributions
from higher partial waves are obvious when $\epsilon\mathtt{>}\Delta$ [the red
dashed curve in Fig. \ref{fig1}(c)].


\begin{figure}[ptb]
\begin{center}
\includegraphics[width=0.6\linewidth]{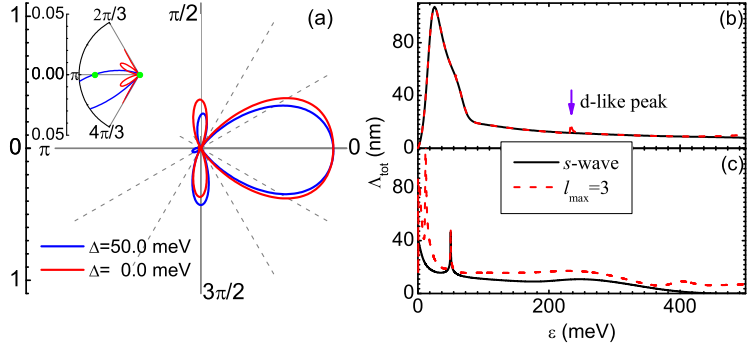}
\end{center}
\caption{ (Color online) The same as in Fig. \ref{fig1} for two identical
scatterers separated by distance $d\mathtt{=}6$ nm on a TI surface.
$\boldsymbol{r}_{1}\mathtt{=}(-3,0)$ and $\boldsymbol{r}_{2}\mathtt{=}(3,0)$.
Other parameters are chosen as the same as in Fig. \ref{fig1}.}%
\label{fig2}%
\end{figure}Now let us discuss the double-scatterer case. The analytical
expressions of the scattering amplitude for multiple scatterers are
complicated, especially when taking into account the higher partial waves and
the band gap. In the limit of $\Delta\mathtt{=}0$, the differential cross
section for two identical scatterers located at $\boldsymbol{r}_{1/2}%
\mathtt{=}\left(  \pm d/2,0\right)  $ becomes symmetric since $M$ is
symmetric. For instance, if we just consider the $s\mathtt{-}$wave, we can
obtain the scattering amplitude%
\begin{align}
f_{0}^{\Delta=0} &  =\sqrt{\frac{2}{i\pi k}}s_{0}\sum_{n,m=1}^{2}e^{i\left(
\boldsymbol{k}\cdot\boldsymbol{r}_{m}-k\hat{r}\cdot\boldsymbol{r}_{n}\right)
}\nonumber\\
&  \times\left\{  \left[  M_{0}^{-1}\right]  _{2n-1,2m-1}\pm i\left[
M_{0}^{-1}\right]  _{2n,2m-1}e^{-i\varphi}\right.  \nonumber\\
&  \left.  \mp i\left[  M_{0}^{-1}\right]  _{2n-1,2m}+\left[  M_{0}%
^{-1}\right]  _{2n,2m}e^{-i\varphi}\right\}  ,\label{23}%
\end{align}
where
\begin{equation}
M_{0}^{-1}=\frac{1}{D}\text{$\left(
\begin{array}
[c]{cccc}%
1 & 0 & s_{0}H_{0} & \pm s_{0}H_{1}\\
0 & 1 & \mp s_{0}H_{1} & s_{0}H_{0}\\
s_{0}H_{0} & \mp s_{0}H_{1} & 1 & 0\\
\pm s_{0}H_{1} & s_{0}H_{0} & 0 & 1
\end{array}
\right)  $}\label{24}%
\end{equation}
with $D\mathtt{=}\sqrt{\det\left\vert M_{0}\right\vert }\mathtt{=}%
1\mathtt{-}s_{0}^{2}\{[H_{0}^{\left(  1\right)  }\left(  kd\right)
]^{2}\mathtt{+}[H_{1}^{\left(  1\right)  }\left(  kd\right)  ]^{2}\}$. It is
easy to find that $f_{0}\left(  \boldsymbol{k},\pi\right)  \mathtt{=}0$, which
indicates that the backscattering is also prohibited. This property can be
clearly seen from the red curve in Fig. \ref{fig2}(a) as well as the inset of
Fig. \ref{fig2}(a). However, the symmetry is reduced by the nonzero gap and
the backscattering can occur, see the blue curve in Fig. \ref{fig2}(a).
Besides, we find that the scattering along the directions of $\varphi
\mathtt{\sim}60^{\circ}$, $120^{\circ}$, $240^{\circ}$, and $300^{\circ}$ is
forbidden since the interference effect along these directions is destructive
during the multiple scattering process. In the forward direction
($\varphi\mathtt{=}0$), however, the interference is still constructive.
Moreover, comparing to the single-scatterer case, the width of resonance peak
is narrowed and the peak is pushed to lower energy due to the interference
during quasiparticle scattering between the impurity pair. Also, the higher
partial waves introduce additional resonance peaks in $\Lambda_{tot}$, see the
red curve in\ Fig. \ref{fig2}(b) and \ref{fig2}(c).

\begin{figure}[ptb]
\begin{center}
\includegraphics[width=0.4\linewidth]{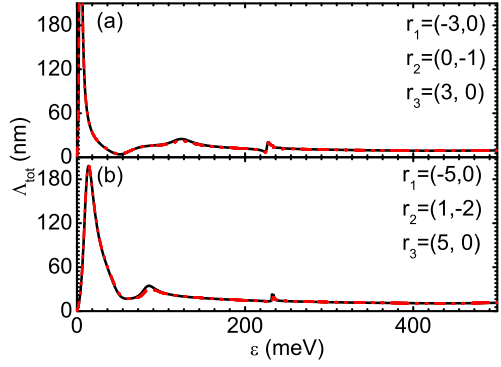}
\end{center}
\caption{ (Color online) Optical theorem Eq. (\ref{20}) for three scatterers.
The black (red dashed) curves are results in the first (second) equality in
Eq. (\ref{20}). $\Delta\mathtt{=}0$ and $l_{\max}\mathtt{=}5$ are chosen in
calculations. }%
\label{fig3}%
\end{figure}

In principle, one can get the cross section as well as other transport
properties for $N\mathtt{\ }$scatterers on the TI surface by the above theory.
Before ending this paper, we calculated $\Lambda_{tot}$ for three scatterers
to demonstrate the optical theorem in Eq. (\ref{20}). For simplicity, we set
$\Delta\mathtt{=}0$ and $l_{\max}\mathtt{=}3$ in calculations, and the results
are shown in Fig. \ref{fig3}, in which the black curve is obtained by
numerical integral while the red dashed curve is obtained from the second
equality in Eq. (\ref{20}). We can see that they are in good agreement with
each other. Therefore we believe the optical theorem in Eq. (\ref{20}) should
be proper to characterize the general multiple scattering process.


In summary, we have developed a general partial-wave multiple scattering
theory for scattering from cylindrically symmetric potentials on TI surfaces
and applied to cross section calculations. We have found that higher partial
waves may introduce some important corrections in low-energy scattering.
Importantly, a resonance peak at the band edge of the gapped TI surface was
found, which should be easily measured in scattering experiments. This
prediction may provide a useful way to determine the band gap of TI surface
states. Our formulation can be extended to magnetic impurity scattering as
well as spin-orbit coupled scattering on TI surface.

This work was supported by NSFC under Grants No. 90921003, No. 60776063, No.
60821061, and 60776061, and by the National Basic Research Program of China
(973 Program) under Grants No. 2009CB929103 and No. G2009CB929300.

\end{document}